\documentstyle[color,12pt]{article}

\def\Tr{{\rm Tr}}
\def\ep{\epsilon}
\def\half{\frac{1}{2}}
\def\quarter{\frac{1}{4}}
\def\fracpow#1#2#3{\frac{#1}{#2}{}_{{}_{\!\!#3}}}
\def\mad{M_{{\rm\scriptscriptstyle\!A\!D}}}
\def\qcd{\Lambda_{\scriptscriptstyle\rm{Q\hspace{-.02cm}C\hspace{-.02cm}D}}}
\def\pg#1#2{P_{\!\!{\scriptscriptstyle#1}\!{\scriptscriptstyle#2}}}
\def\sym{S_{{\rm Y\!M}}}
\definecolor{gray}{rgb}{.4,.4,.4}
\def\cinza#1{\textcolor{gray}{#1}}
\begin{document}

\thispagestyle{empty}
\begin{flushright}
{\tt %hep-th/01xxxxxxx\\
\cinza{DIAS-STP-01-17}\\}
\end{flushright}
\vspace{.2cm}
\begin{center}
  {\large \bf Abelian projected $SU(2)$ Yang-Mills action\\
  for renormalisation group flows}\\[5ex]
 
  {Filipe Freire}\,\footnote{E-Mail:\ filf@stp.dias.ie}\\[2ex]

  School of Theoretical Physics,\\ Dublin Institute for Advanced Studies,\\
  10 Burlington Rd, Dublin 4, Ireland\\and\\Department of Mathematical
  Physics,\\ National University of Ireland,\\ Maynooth, Ireland\\[4ex]

  {\small \bf Abstract}\\[2ex]
\begin{minipage}{14cm}{\small
    The dual Meissner effect scenario of confinement is discussed
    by studying the low energy regime of $SU(2)$
    Yang-Mills in a maximal Abelian gauge. The Abelian projected
    effective action is computed perturbatively. This serves
    as an input for a study of the non-perturbative regime, which is
    undertaken using exact renormalisation group methods. It is argued
    that the effective action derived here contains the relevant
    degrees of freedom for confinement if ultraviolet
    irrelevant vertices are retained.\\[4ex] 

    }
\end{minipage}
\end{center}
\newpage \pagestyle{plain} \setcounter{page}{1}

\renewcommand{\thefootnote}{\arabic{footnote}}
\setcounter{footnote}{0}

\noindent
{\bf Introduction}\\[1ex]
The understanding of the low energy behaviour of QCD is one of the
long standing challenges in theoretical particle physics.  In this
letter we elaborate on the dual Meissner mechanism for
confinement \cite{Nambu:1974zg} which is best studied in maximal Abelian
gauges (MAGs). In these gauges the condensation of colour monopoles
is explicit \cite{Mandelstam:1976pi,'tHooft:1981ht}. 't\,Hooft
\cite{'tHooft:1981ht} conjectured that these $U(1)$-projected
singularities govern confinement which sets the scenario of Abelian
dominance.

This conjecture has been tested on the lattice. In gauge fixed lattice
$SU(2)$ and $SU(3)$ strong evidence for Abelian dominance has been
found \cite{Suzuki:1990gp}. It has been also shown that monopole
condensates alone account for most of the string tension in the
confining phase \cite{Smit:1994vt}. Other studies favour vortices as
the relevant degrees of freedom \cite{Engelhardt:2000wr}. Preliminary
lattice results for the scale of Abelian dominance give $\mad\simeq
1.2$ GeV, clearly indicating that $\mad > \qcd$ \cite{Amemiya:1999zf}. 
Analytical aspects of Abelian dominance have been studied as
well. This includes phenomenological models \cite{Ezawa:1982bf} and 
perturbative studies \cite{Kondo:1998pc}. Moreover, a mass
generating mechanism involving ghost pairs has been discussed by
introducing a parallel with BCS superconductivity
\cite{Schaden:1999ew,Kondo:2000ey}.

The scale separation allows for a
perturbative computation of the effective action $\Gamma_\Lambda$ at
scales $\Lambda\gg\mad$. However, non-perturbative methods are needed at
scales smaller than $\mad$. The exact renormalisation group (ERG) for
gauge theories is such a method \cite{Reuter:1994kw,Litim:1998qi}.
The ERG requires the effective action $\Gamma_\Lambda$ as a key
input and provides the ideal framework for investigating the above
scenario.  Its viability in the present context has already been
highlighted in \cite{Ellwanger:1998wv}. So far, the ERG has
been formulated for general linear gauges \cite{Litim:1998nf}.
However, the extension to the non-linear maximal Abelian gauges poses
no new problems. One can easily transfer the result of studies carried
out in the background field formalism \cite{Freire:1996db}.

In this letter, we compute an Abelian effective action for $SU(2)$ YM
in a MAG. This is done by integrating over the charged gauge fields in
the $SU(2)/U(1)$ coset broken by the MAG condition. All vertices up to
the first UV non-relevant ones are kept. This provides us with an 
initial effective action for the ERG flow. We argue that this initial
effective action contains the prerequisites for the dual Meissner effect: 
a coupling between a dual Abelian gauge field and a monopole
current, and a kinetic term for the auxiliary tensor field. Finally,
under certain conditions, we show the ERG flow drives the system into
the confining phase.

\vspace{.25cm}
\noindent {\bf The SU(2) action}\\[1ex]
Let us begin by expressing the $SU(2)$ YM action in a four dimensional
Euclidean space in terms of a neutral and charged vector fields. In
conventional variables the action is
\begin{equation}
  \label{ymaction}
  \sym = \frac{1}{4}\int_x F_{\mu\nu}^a\,F_{\mu\nu}^a\,,
\end{equation}
where $\int_x$ is a short hand for $\int d^4x$ and the field strength is
$F_{\mu\nu}^{a}=\partial_\mu A^a_\nu - \partial_\nu A^a_\mu +
  g\,\ep^{abc}A^b_\mu A^c_\nu\,$, with $g$ the gauge
coupling parameter and $\ep^{abc}$ the canonical antisymmetric
$3$-tensor. We now introduce the new field variables
\begin{equation}
  \label{newvariables}
  A_\mu=A_\mu^3\,,\hspace{1cm}\phi_\mu = \frac{1}{\sqrt2}\, 
  (A_\mu^1-iA_\mu^2)\,,\hspace{1cm}\phi^\dagger_\mu =
  \frac{1}{\sqrt2}\,(A_\mu^1+iA_\mu^2)\,.
\end{equation}
These are easily recognised as the gauge field components associated
with the diagonal $SU(2)$ generator and the off-diagonal lowering and
raising ones respectively.

When expressed in terms of the fields~(\ref{newvariables}) the YM
action takes the form
\begin{equation}
  \label{newymaction}
  \sym = \frac{1}{4} \int_x(f_{\mu\nu}+C_{\mu\nu})\,
  (f_{\mu\nu}+C_{\mu\nu}) +\half \int_x \Phi^\dagger_{\mu\nu}
  \Phi_{\mu\nu}\,,
\end{equation}
where $f_{\mu\nu}$ and $\Phi_{\mu\nu}$ are the field strengths for
the Abelian and the charged vector fields, respectively, given by
$f_{\mu\nu} = \partial_\mu A_\nu - \partial_\nu A_\mu\,$ and
$\Phi_{\mu\nu} = D_\mu \phi_\nu - D_\nu \phi_\mu\,$,
with $D_\mu = \partial_\mu -igA_\mu$ the Abelian covariant derivative.
Moreover, $C_{\mu\nu}$ in~(\ref{newymaction}) is a real,
antisymmetric, quadratic combination of the off-diagonal fields,
\begin{equation}
  \label{ctensor}
  C_{\mu\nu} = ig\left(\phi_\mu^\dagger\ \phi_\nu -
  \phi_\nu^\dagger\ \phi_\mu\right).
\end{equation}

The YM action~(\ref{newymaction}) is explicitly invariant under a
$U(1)$ subgroup of $SU(2)$, where under $U(1)$ the charged vector
fields are transformed only by an Abelian phase factor. Of course
the action is $SU(2)$ invariant as well.
There are two types of quartic coupling in~(\ref{newymaction}).
Those between the neutral and the charged vector fields,
$A_\mu A_\mu \phi^\dagger_\nu \phi_\nu - A_\mu A_\nu
  \phi^\dagger_\mu \phi_\nu$\,,
and the charged vector fields self-couplings,
\begin{equation}
 \label{quartic}
  \quarter\ C_{\mu\nu} C_{\mu\nu}=-\half\ g^2
  \left(\phi_\mu^\dagger\ \phi^{}_\nu\ \phi_\mu^\dagger\ \phi^{}_\nu -
    \phi_\mu^\dagger\ \phi^{}_\nu\ \phi_\nu^\dagger\ \phi_\mu\right).
\end{equation}
Hitherto, we have restricted the discussion to the classical level.  The
quantisation of YM theories is non-trivial because of the
constraints.  Besides, if we insist on a covariant formulation
we need to handle unphysical zero modes.
It is well known how to treat this problem.  Here we pursue the
path integral quantisation with gauge fixing and ghost fields.

The action~(\ref{newymaction}) is well-suited to study 
Abelian dominance in the continuum because of its explicit
$U(1)$ invariance.  For the gauge-fixing we choose a MAG condition
which leaves the $U(1)$ invariance in~(\ref{newymaction}) unbroken. Then, 
the remnant gauge freedom is fixed with a Lorentz condition,
i.e. respectively
\begin{eqnarray}
  \label{eq:mag}
  F^\pm[\phi,\,A] := (\partial_\mu \pm ig A_\mu)\phi_\mu = 0\,,
  \hspace{.5cm}F[A] := \partial_\mu A_\mu = 0\,.
\end{eqnarray}
Though Gribov copies exist in a MAG they do not give rise to any
sizable effects \cite{Hart:1997ar} and can be ignored. Then the gauge
fixing sector including the ghost action is
\begin{eqnarray}
  \label{gaugefixing}
  S_{{\rm g}} &=& \frac{1}{2\xi}\int_x \left(\partial_\mu A_\mu\right)^2
  + \frac{1}{\xi^\prime}\int_x \left(D_\mu \phi_\mu\right)^\dagger
  \left(D_\nu \phi_\nu\right) - \int_x \bar c_+\left(D_\mu^\dagger
  D_\mu^\dagger + g^2 \phi_\mu^\dagger \phi_\mu\right)c_+ -\nonumber\\[.5cm] 
  & &- \int_x \bar c_-\left(D_\mu D_\mu + g^2 \phi_\mu^\dagger
  \phi_\mu\right)c_- +\ g^2\!\int_x \bar c_+ c_-\ \phi_\mu^\dagger
  \phi_\mu^\dagger +\ g^2\!\int_x \bar c_- c_+\ \phi_\mu \phi_\mu\,.
\end{eqnarray}
The quantum theory for the gauge fixed action $S = \sym\!+ S_{{\rm g}}$
will now be studied in a path integral representation.

\vspace{.25cm}
\noindent {\bf Abelian dominance}\\[1ex]
The suppression of the $\phi_\mu$ fields \cite{Suzuki:1990gp}
suggests that they acquire a
mass \cite{Amemiya:1999zf,Schaden:1999ew,Kondo:2000ey} dynamically, 
but this mass cannot be computed perturbatively due to
BRS invariance. We expect that this mass sets the
scale of Abelian dominance,
$M_{\!{}_{\rm{A\!D}}}$, below which a qualitative change of the
relevant dynamical variables takes place. At a
much lower scale this will lead to confinement. Preliminary
results from the lattice give $\mad\!\simeq\!1.2$ GeV
\cite{Amemiya:1999zf}. Though we expect that this value might decrease, 
or that for $SU(3)$ it should
be smaller, it is an indication that the scale of Abelian dominance is
larger than the confining scale $\qcd$.

The first step towards an effective Abelian theory is to integrate
over the charged vector fields. The presence of vertices with four
of these fields~(\ref{quartic}) hinders a straightforward integration.
This problem can be surmounted with the introduction of an auxiliary
tensor field, $B_{\mu\nu}$.

The tensor field is introduced in~(\ref{newymaction}) via the replacement,
\begin{eqnarray}
  \label{tensor}
  \quarter \int_x C_{\mu\nu} C_{\mu\nu}\ \longrightarrow\ - \quarter
  \int_x \tilde B_{\mu\nu} \tilde B_{\mu\nu} + \half \int_x \tilde
  B_{\mu\nu} C_{\mu\nu}.
\end{eqnarray}
It follows that the equation of motion for $B_{\mu\nu}$ is
$\tilde B_{\mu\nu} = C_{\mu\nu}$. After 
inserting it back into~(\ref{tensor}), the term
quadratic in $C_{\mu\nu}$ is recovered. When performing the
replacement~(\ref{tensor}) in the full action $S$ the integration
over the charged vector fields becomes Gaussian. Now we focus our attention
on the part of the action quadratic in these fields. We get 
\begin{equation}
  \label{eq:matrix}
  S_{\phi^2} = \int_x \left(\phi_\mu^\dagger\ 
  a_{\mu\nu}^{-+}\ \phi_\nu + \phi_\mu\ a_{\mu\nu}^{+-}\ 
  \phi_\nu^\dagger + \phi_\mu\ a_{\mu\nu}^{++}\ \phi_\nu +
  \phi_\mu^\dagger\ a_{\mu\nu}^{--}\ \phi_\nu^\dagger\right),
\end{equation}
where the elements of the rank two matrix $a_{\mu\nu}^{SS'}$, with 
${\scriptstyle S}, {\scriptstyle S'} = {\scriptstyle\pm}$\,, are
\begin{eqnarray}
  \label{eq:matrixelements}
  a_{\mu\nu}^{-+}\!\!\!&=&\!\!- \half\ g_{\mu\nu} \left(D_{\rho}
  D_{\rho} + g^2\ \bar c_+ c_+\right) + \frac{1}{2\xi^\prime}
  (\xi^\prime - 1)D_{\mu} D_{\nu} + \frac{ig}{2\xi^\prime} 
    \left[(\xi^\prime + 1)f_{\mu\nu} +\xi^\prime
    \tilde B_{\mu\nu}\right],\nonumber\\[.3cm]
    a_{\mu\nu}^{+-}\!\!\!&=&\!\!\left[a_{\mu\nu}^{-+}\right]^\dagger\!\!,
    \hspace{1cm}a_{\mu\nu}^{++} = g^2\,g_{\mu\nu}\, \bar c_- c_+\,,
    \hspace{1cm}a_{\mu\nu}^{--} = \,\left[a_{\mu\nu}^{++}\right]^\dagger\!\!.
\end{eqnarray}
From now on, for simplicity,  we take $\xi^\prime = 1$.

Due to the Gaussian integration the effective action receives a
contribution of the form, $\half\,{\rm Tr} \ln\,a_{\mu\nu}^{SS'}$. In 
a Schwinger proper-time representation it reads 
\begin{eqnarray}
  \label{schwingerrep}
  \half\,{\rm Tr}\ln \left(a_{\mu\nu}^{SS'}\right) = -\half\ 
  \lim_{s\to0}\ \frac{d}{ds}\left(
  \frac{\mu^{2s}}{\Gamma(s)}\int_0^\infty\!\!dt\ t^{s-1}\ {\rm
    Tr}\left[\ e^{-t a_{\mu\nu}^{SS'}} - e^{-t
    {a_o}_{\mu\nu}^{SS'}}\ \right]\right),
\end{eqnarray}
where ${a_o}_{\mu\nu}^{SS'}$ is $a_{\mu\nu}^{SS'}$ at vanishing fields
and $a_0$ is introduced as a regulator. The trace {\rm Tr} is over covariant
and group indices and a complete set of states. Finally, the constant
$\mu$ is an arbitrary regulator scale.
The part of the trace corresponding to the sum over a complete set of states
takes a very convenient integral form for a complete base of plane waves, 
which will be very suitable to develop systematic approximations.
Together with the change of variable, $k_\rho \rightarrow \sqrt{t}\,k_\rho$\,,
the trace written in a base of plane waves is now
\begin{eqnarray}
    \label{trace}
{\rm Tr}\left[e^{-t a_{\mu\nu}^{SS'}}\right] =
  \fracpow{\,\,t^{-2}}{(2\pi)}{4} \int_x \int_k e^{-\half k^2} {\rm tr}
  \left[\exp\Big[{\scriptstyle{-\,t\,a_{\mu\nu}^{SS'} +\, i\,\sqrt{t}\,\,
  g_{\mu\nu}\left(\,\bar\delta^{S+}\bar\delta^{S'\!\!-}\,k_\rho D_\rho\, 
  +\,\bar\delta^{S-}\bar\delta^{S'\!\!+}\,k_\rho D^\dagger_\rho\right)}}
  \Big]\right],
\end{eqnarray}
where {\rm tr} is the trace restricted to the covariant and group
indices. We have also introduced the inverse Kronecker delta,
$\bar\delta^{SS'} = 1$ if ${\scriptstyle S} \neq {\scriptstyle S'}$ and
$\bar\delta^{SS'} = 0$ if ${\scriptstyle S} = {\scriptstyle S'}$.
After inserting~(\ref{trace}) into the right-hand side of~(\ref{schwingerrep})
and Taylor expanding the exponentials, the contribution to the effective
action becomes 
\begin{eqnarray}
  \label{trace4}
  \half\,{\rm Tr}\ln \left(a_{\mu\nu}^{SS'}\right) =
  - \half\lim_{s \to0}\ \frac{d}{ds}\
   \frac{\mu^{2s}}{\Gamma(s)}\int_0^\infty\!\!dt\ t^{s-3}\ \int_x\
  \!\sum_{n=0}^{\infty}\frac{(-1)}{n!}^n\!\!\int_k e^{-\half k^2}
   \!\times\hspace{1cm}\nonumber\\[.25cm]
   {\rm tr}
  \left[\left({\scriptstyle{t\,a_{\mu\nu}^{SS'} - i\,\sqrt{t}\,
  g_{\mu\nu}\left(\,\bar\delta^{S+}\bar\delta^{S'\!\!-}\,k_\rho D_\rho 
  +\,\bar\delta^{S-}\bar\delta^{S'\!\!+}\,k_\rho
  D^\dagger_\rho\right)}}\right)^n\!\!-
 \left({\scriptstyle{t\,{a^{}_0}_{\mu\nu}^{SS'} - i\,\sqrt{t}\,
  g_{\mu\nu}\,\bar\delta^{SS'}\,k_\rho \partial_\rho
  }}\right)^n\,\right].\hspace{-1cm}
\end{eqnarray}
In order to progress beyond Eq.~(\ref{trace4}), it is necessary to
employ approximations. The momentum integration is convergent but not
solvable. We will proceed with an
expansion on small $t$ which provides a systematic approach of computing
the effective vertices for short range interactions
by decreasing importance in their UV relevance.

\vspace{.25cm}
\noindent {\bf Ultraviolet relevance}\\[1ex]
On the right-hand side of~(\ref{trace4}) only the integer powers of 
the Schwinger proper-time do not vanish after the integration over
the momenta, because the integrand for the half integer powers of $t$
is odd in the momenta.  The integration in the Schwinger proper-time
is now splitted into two part,
$\int_{_0}^{1/\Lambda^{\!2}}\!+\int_{1/\Lambda^{\!2}}^{^\infty}$\ . 
The scale $\Lambda$ is a UV scale larger than $\mad$ chosen so that the
integration for $t < 1/\Lambda^{\!2}$ is negligible due to Abelian dominance.
Therefore, we keep only the second part of the integration with 
$1/\Lambda^{\!2}$ in the lower bound. 
Since we are interested in the leading UV vertices
only the contributions coming from this bound are retained.
After the integration, the expansion in powers of $t$ emerges
as an expansion in vertex operators of decreasing UV relevance. Here
we keep the terms of this expansion up to the first non-relevant
ones, i.e. ${\cal O}(1/\Lambda^{2})$. The resulting effective
action, with $\mu=\Lambda$, is
\begin{eqnarray}
  \label{adea}
  S_{\rm{eff}}&=&\quarter\left(1+g^2\fracpow{5\gamma}{12\pi}{2}\right)
  \int_x f_{\mu\nu}f_{\mu\nu} +
  \frac14 \int_x B_{\mu\nu} \left( \fracpow{g^2}{96\pi}{2}\,
  \fracpow{\Box}{\Lambda}{2} -
  1 + g^2\fracpow{\gamma}{8\pi}{2} \right)\,B_{\mu\nu} +\nonumber\\[.5cm]
  & &\!+\ g^2\fracpow{\gamma}{8\pi}{2}\,\!\int_x \tilde
  f_{\mu\nu}\,B_{\mu\nu} - \int_x \bar c_+\left(D_\mu^\dagger D_\mu^\dagger
  + \fracpow{g^2}{2\pi}{2}\Lambda^2\right) c_+ 
  -\int_x \bar c_-\left(D_\mu D_\mu +
  \fracpow{g^2}{2\pi}{2}\Lambda^2\right)c_- - \nonumber\\[.5cm]
  &&\!-\,g^4\fracpow{\gamma}{2\pi}{2}\int_x \bar c_+ c_+ \bar c_- c_-
   + \frac{1}{2\xi}\int_x
  \left(\partial_\mu A_\mu\right)^2 +\nonumber\\[.5cm]
  & &\!+ \fracpow{1}{\Lambda}{2}\left(\, \fracpow{g^2}{96\pi}{2}
  \int_x \tilde f_{\mu\nu}\Box B_{\mu\nu} +
  \fracpow{11\,g^2}{960\pi}{2}\int_x
  f_{\mu\nu}\Box f_{\mu\nu} +
  \fracpow{g^4}{192\pi}{2}\int_x \left(\bar c_+ c_+\!+ \bar c_- c_-\right)
  B_{\mu\nu}B_{\mu\nu}\right. +\nonumber\\[.5cm]
  & &\!+ \left.\fracpow{g^4}{48\pi}{2} \int_x \left(\bar c_+ c_+\!+
  \bar c_- c_-\right)  
  \tilde f_{\mu\nu}B_{\mu\nu} + \fracpow{g^4}{96\pi}{2}
  \int_x \left(\bar c_+ c_+\!+ \bar c_- c_-\right) 
  f_{\mu\nu}f_{\mu\nu}\right.+\nonumber\\[.5cm] 
  & &\!\left.+ \fracpow{g^4}{48\pi}{2} \int_x (\bar c_- c_+)
  D^{(2)}_\mu D^{(2)}_\mu\,(\bar c_+ c_-)\right) + 
  {\cal O}\left(\fracpow{1}{\Lambda}{4},\,B^4\right),
\end{eqnarray}
with $D^{(2)}_\mu = \partial_\mu - 2ig A_\mu$\,, the covariant derivative 
for a charged two field. Note that the ${\cal O}(\Lambda^{2})$ terms 
correspond to a mass renormalisation, and the coupling and wave function 
renormalisation are identified in~(\ref{adea}) as the terms proportional 
to the Euler gamma. The latter is proportional to $\left(\ln[\Lambda/\mu] 
- \gamma/2\right)$ but we have set the renormalisation constant scale 
$\mu$ in~(\ref{schwingerrep}) equal to $\Lambda$.  We note that in a
more complete treatment where the hard modes of the remaining fields
are integrated out to one-loop order down to $\Lambda$,
the ${\cal O}(\Lambda^{2})$ ghost terms are cancelled in accordance to BRS
invariance. Finally, as expected, the effective action~(\ref{adea}) is
$U(1)$ invariant.

The UV marginal terms account for wave function and coupling renormalisations.
There are two new vertices that were not present at tree level: the
coupling between the dual of the Abelian field 
strength and the tensor field, and the 4-ghosts vertex, the first
terms in the second and third line of~(\ref{adea}) respectively. 
As far as monopoles are concerned, the
relevant vertex appearing at this order is the one involving the
auxiliary tensor field $B_{\mu\nu}$. Kondo \cite{Kondo:1998pc} has shown by
using a Hodge decomposition of $B_{\mu\nu}$ that the
$\tilde f_{\mu\nu}\,B_{\mu\nu}$ term
encapsulates the coupling between a gauge field potential and a magnetic
current $J^{M}_\nu=\partial_\mu \tilde f_{\mu\nu}$.
Note, that no term involving $f_{\mu\nu}\,B_{\mu\nu}$ is generated at
this order.  At present, we do not have a clear explanation for it, but
its existence seems to imply that instead of a condensate of monopoles,
we might have a condensate of dyons.

Finally, we comment on the first correction of UV irrelevant
vertices. One vertex was singled out in~(\ref{adea}), namely the term
governing the dynamics of the tensor field. This term was placed in
the first line of the right-hand side of~(\ref{adea}). The remaining
${\cal O}(1/\Lambda^{2})$ terms are displayed in the last three lines
of~(\ref{adea}). The irrelevant vertex, included in the
leading UV terms, corresponds to the leading ghost free term in lowest
order in a derivative expansion. Below, we will discuss this operator
at greater length.

A few remarks on the scales $\mu$ and $\Lambda$ should be made at this stage.
The scale $\mu$ is the renormalisation group scale coming from the
dimensional regularisation in the Schwinger proper-time
representation~(\ref{schwingerrep}).  In order to
define the effective theory a UV scale $\Lambda$ is introduced and the
resulting one-loop logarithms are of the form $\ln[\Lambda/\mu]$ as
referred above.  The choice $\mu=\Lambda$ is naturally a convenient one.
The resulting effective action,~(\ref{adea}), is in the spirit of
\cite{Weinberg:1980wa} an appropriate initial condition at the scale
$\Lambda$ for an ERG analysis of the low energy theory.

Information on the relative magnitudes of the scale $\Lambda$ to $\mad$
is necessary to complete this discussion.  The sole requirement so far
on $\Lambda$ was that it is in the perturbative region of $SU(2)$ YM.
This ensures that the coefficients of the different vertices can be
determined reliably about this UV scale.
However, in the present problem we expect two more, though related,
scales: the confinement scale $\qcd$ and the scale of Abelian
dominance $\mad$, with $\mad > \qcd$.
In order to study how the presence of $\mad$ might condition the choice
of $\Lambda$ we have also calculated the effective action when
the charged vector fields have a mass $\mad$ inserted
in~(\ref{eq:matrixelements}).
We found that the coefficients in~(\ref{adea}) receive corrections in
a power series with respect to $\mad^{\,2}/\Lambda^2$. This indicates
that in order for~(\ref{adea}) to provide a reliable effective action
we need to require $\Lambda \gg \mad$. Therefore, we can safely expect
to improve the computation of~(\ref{adea}) within the presently
available methods.

\vspace{.25cm}
\noindent {\bf Exact renormalisation group and confinement}\\[1ex]
The main goal here is to show that
qualitatively the Abelian effective action~(\ref{adea}) contains 
the relevant degrees of freedom for confinement.
Therefore, in the following we ignore the ghost sector and only keep
terms up to quadratic order in a derivative expansion.  The effective
action is used as the initial condition to the ERG equations at the
scale $\Lambda$.

The ERG equations are flow equations for the effective action
$\Gamma_k$ with respect to the scale $k$ running from $k=\Lambda$ to
$k=0$. The $k$ dependence comes from the insertion of IR cut-off
functions $R_k$ in $\Gamma_k$ (for the gauge and ghost fields in the
present case). The flow equation has the structure of a one-loop equation
\begin{eqnarray}
  \label{flow-eqn}
  \partial_t \Gamma_k = \half\, \Tr\left[{\partial_t R_k}\,
  \left(\Gamma_k^{(2)} + R_k\right)^{-1}\right],\hspace{.2cm}t=\ln k\,,
\end{eqnarray}
where $\Tr$ stands for the sum over all fields and group indices as well
as the integration over configuration (or momentum) space, whilst
$\Gamma^{(2)}_k$ is the fully dressed 1PI functional \cite{Reuter:1994kw}.
Diagrammatically, the right-hand side of Eq.~(\ref{flow-eqn}) is the
sum of one-loop diagrams with a $\partial_t R_k$ insertion and full
propagators.

For the problem discussed in this letter, we
have seen that the Abelian effective action~(\ref{adea}) contains the
leading and near-to-leading UV operators. Therefore, for the 
effective action $\Gamma_k$ we choose as an Ansatz,
\begin{eqnarray}
  \label{ansatz}
  \Gamma_k = \int_x\!\Big\{\,\frac{Z_{\!A}}{4}\,f_{\mu\nu}f_{\mu\nu} +
  \frac{1}{4}\,(-Z_{\!B}\,\Box + M_{\!B}^2)\,B_{\mu\nu}B_{\mu\nu} +
  \frac{Y}{2}\,\tilde f_{\mu\nu} B_{\mu\nu}
  +\frac{1}{2\xi}\,\left(\partial_\mu A_\mu \right)^2\!+
  \textrm{ghosts} \Big\}.
\end{eqnarray}
The coefficients $Z_{\!A}$, $Z_{\!B}$, $M_{\!B}^2$ and $Y$ are
renormalisation functions which depend on $k$.
The vertices in this Ansatz consist of the first order terms
in~(\ref{adea}) plus the leading ${\cal{O}}(1/\Lambda^{2})$ term
in a derivative expansion. The dynamics of the Abelian gauge field and
the tensor field are governed by the first two terms respectively.
The third term establishes the coupling between these two fields.
Effectively it couples the magnetic current to a dual magnetic
field \cite{Kondo:1998pc}. The fourth term is the gauge fixing and
``ghosts'' stands for all the terms involving ghosts up to the first
order in derivative expansion. These terms are
left out as their presence would not have altered 
the results presented below. In any case, as we have an
effective Abelian theory, they should not play a major role for 
the leading structure of the gauge field propagator.

Next, we show that this Ansatz contains enough information to
exhibit a confining phase at low energies. As a signature of
confinement we search for a $1/p^4$ singularity in the gauge field
propagator. It is well established that a propagator with this feature
leads to an area law in a Wilson loop \cite{West:1982bt}.
By implementing the Ansatz~(\ref{ansatz}) the problem of finding the
flow of the functional $\Gamma_k$ is reduced to that of determining
the running of $Z_{\!A}$, $Z_{\!B}$, $M_{\!B}^2$ and $Y$ as functions of $k$.

The initial condition for the ERG flow is assigned by taking
$\Gamma_{k=\Lambda} \approx S_{\rm eff}$ which, in particular,
for the new coefficient $Z_{\!B}$ reduces to $Z_{\!B}(\Lambda) \approx 0$. 
Of course, this simply reflects that $B_{\mu\nu}$ was introduced as an
auxiliary field.  From $S_{\rm eff}$ in~(\ref{adea}) we have that
$M_{\!B}^2 < 0$ at the initial scale. 
However the coefficient of any vertex of 
higher power in the tensor field $B_{\mu\nu}$ can be shown to be positive.
It seems to indicate that $B_{\mu\nu}$ acquires a VEV but most importantly 
it guarantees the stability of the effective action along the tensor field 
direction.  Therefore, by taking $B_{\mu\nu}$ to be the fluctuation
field about its global minimum we guarantee $M_{\!B}^2$
to be positive. This will suffice for our present proposes.

A useful feature of working with an Ansatz as~(\ref{ansatz}) is that
propagators can be expressed as functionals of the renormalisation
functions for any value of $k \leq \Lambda$\,. The functional form of
the propagators is determined by simply inverting the two point 1PI
Green's functions. Then, the gauge field propagator of 
$\Gamma_k$ as given in (\ref{ansatz}), is
\begin{equation}
  \label{eq:propagators}
  \left(\pg{A}A\right)_{\mu\nu} =
  \left(\delta_{\mu\nu} - \frac{p_\mu p_\nu}{p^2}\right)
  \frac{Z_{\!B}\,p^2 + M_{\!B}^2}{Z_{\!A}\,Z_{\!B}\,p^4 +
  p^2\,(M_{\!B}^2\,Z_{\!A} - Y^2)} +
  \xi\ \frac{p_\mu p_\nu}{p^4}\,.
\end{equation}
In the Landau gauge, $\xi=0$, we observe that the denominator of the
gauge field propagator~(\ref{eq:propagators}) is dominated by the
$p^4$ term if
\begin{equation}
  \label{fixedpoint}
  M_{\!B}^2(k)\,Z_{\!A}(k) = Y(k)^2
\end{equation}
is an IR stable quasi-fixed point.
Then, if the condition~(\ref{fixedpoint}) is realised, the gauge field
propagator will have a $1/p^4$ behaviour at small momentum, if the $p$
dependence on the numerator becomes negligible, i.e.
\begin{eqnarray}
  \label{eq:qcd}
  Z_{\!B}(k_c)\,p^2 \ll M_{\!B}^2(k_c) \ \ \ \Rightarrow \ \ \ \sqrt{p^2} \ll
  \frac{M_{\!B}(k_c)}{\sqrt{Z_{\!B}(k_c)}}:= \qcd\,.
\end{eqnarray}
This result should be discussed in parallel to the findings of Ellwanger
\cite{Ellwanger:1998wv}. 
Our treatment differs in three main aspects: (a) we work
in a MAG instead of the Landau gauge; (b) we present a systematic
computation of the action on which we base our Ansatz.
Ellwanger's Ansatz is based on the requirement of full BRS invariance
and the Abelian projection follows as a truncation to a diagonal Abelian
component. Our systematic calculation provides a more direct link with
the parameters of the theory; (c) the tensor field used by Ellwanger
is the dual of the entire field strength of the 't\,Hooft-Polyakov monopole
\cite{'tHooft:1974qc} while we consider only the dual of its quadratic part.

In his work, Ellwanger found a condition
analogous in form to~(\ref{fixedpoint}). Furthermore, evidence
for a quasi-fixed point was found within his approximations. 
Further investigation is required to verify, whether the same applies
in the present case. However, in view of the similarities
we expect an analogous outcome. More recently, Ellwanger and Wschebor
\cite{Ellwanger:2001yy}, have found that for an effective gauge theory where
it is assumed that the charged fields as well as the ghosts have been
integrated over, the equivalent to condition~(\ref{fixedpoint}) is
relaxed to an inequality which in the present case would read
$M_{\!B}^2(k)\,Z_{\!A}(k) - Y(k)^2 < 0$, for $k>0$.
The equality is recovered in the IR limit $k\to0$.

\vspace{.25cm}
\noindent {\bf Confinement and the dynamics of the $B_{\mu\nu}$ field}\\[1ex]
We have seen that the propagator of the gauge field can, under
certain conditions, lead to a linear
effective quark potential. The IR $1/p^4$ singularity occurs because:
(a) there is a term proportional to $p^4$ in the denominator of the
propagator that becomes prominent when the condition~(\ref{fixedpoint}) 
holds; (b) at small momentum, when~(\ref{eq:qcd}) is fulfilled, the
numerator becomes $p^2$ independent. The proportionality factor to
$p^4$ in the denominator is $Z_{\!A} Z_{\!B}$ where $Z_{\!A}$ is
always non-zero as the $U(1)$ gauge field remains dynamical in any
phase.  However, the situation is different for $Z_{\!B}$. 
The tensor field is not dynamical at short distances,
$Z_{\!B} \approx 0$, which reflects the UV irrelevance of the
$B_{\mu\nu}$ field kinetic term. 

Therefore in order to have a non-vanishing factor $Z_{\!A} Z_{\!B}
\not\approx 0$, somewhere in between the
deep UV region $k\simeq\Lambda$ and the confinement scale $\qcd$,
the tensor field kinetic vertex must undergo a crossover that will
make it relevant in the IR. We expect this crossover scale to be
linked to the scale of Abelian dominance $\mad$. Above $\mad$ the dynamics of
$B_{\mu\nu}$ is protected by the still unsuppressed off-diagonal gauge
fields by a ${\cal{O}}(1/\Lambda^{2})$ factor. Below $\mad$ the effects of 
gauge fields
associated with the off-diagonal components quickly loose prominence. This
is counterbalanced by terms involving
the $U(1)$ invariant tensor field. Consequently, $Z_{\!B}$ plays an
equally relevant role in the Abelian dominated regime when compared
with the other renormalisation functions. Hence, we can conclude
from~(\ref{eq:propagators}) that the scenario described above leading
to a $1/p^4$ behaviour is plausible. Polonyi's view of confinement as
an irrelevant-to-relevant crossover \cite{Polonyi:1995jz} of some UV
irrelevant operator is in line with the present scenario.

\vspace{.25cm}
\noindent {\bf Summary and discussion}\\[1ex]
We have combined the benefits of working in a MAG and using ERG
methods to study the monopole mechanism for confinement. The present
approach is based on the assumption that an intermediate Abelian
dominance scale $\mad$ is dynamically generated, a viewpoint supported
by lattice results.  Starting from a pure
$SU(2)$ YM theory an Abelian effective action was derived. In the
calculation we used a maximal Abelian gauge, introduced an auxiliary
tensor field, and integrated over the gauge fields in $SU(2)/U(1)$.
This serves as an initial effective action for an ERG flow.  The
initial UV scale $\Lambda$ is well inside the perturbative region of
the theory.  It is argued that confinement occurs if the ERG flow at
low energies settles about an IR stable quasi-fixed point. Here,
confinement hinges on an irrelevant-to-relevant crossover for the
kinetic term of the auxiliary tensor field. We expect the 
crossover scale to be the scale of Abelian dominance $\mad$.

\vspace{.5cm}

\noindent
{\bf Acknowledgements}\\[1ex]
I thank Ulrich Ellwanger, Daniel Litim, Tim Morris, Jan Pawlowski and 
Nicolas Wschebor for helpful discussions. The good working
environment provided by the Physics Department at Southampton
University and the PPARC financial support at the early stages of
this work are also acknowledged.

\end{document}